\documentclass[a4paper, 9pt, conference]{IEEEtran}

\makeatletter
\IEEEoverridecommandlockouts
\usepackage{cite}
\usepackage{amsmath,amssymb,amsfonts}
\usepackage{algorithmic}
\usepackage[]{graphicx}
\graphicspath{{Figures/}}
\usepackage{textcomp}
\usepackage{lipsum}
\usepackage{xcolor}
\usepackage{multirow}
\usepackage{balance}
\usepackage{url}

\usepackage{amsmath,url}

\usepackage{cite}
\usepackage{amsmath,amssymb,amsfonts}

\usepackage{subfig}
\usepackage{lipsum}
\usepackage{balance}
\usepackage{textcomp}
\usepackage{xcolor}
\usepackage{multirow}
\usepackage{color}
\usepackage{xcolor}
\usepackage{algorithm} 
\usepackage{algorithmic}
\usepackage{listings}
\usepackage{courier}
\usepackage{float}

\usepackage{tikz}
\usepackage{lipsum}

\usepackage{comment}

\usepackage[left=0.5in, right=0.5in, top=1in, bottom=1in]{geometry}

\def\BibTeX{{\rm B\kern-.05em{\sc i\kern-.025em b}\kern-.08em
    T\kern-.1667em\lower.7ex\hbox{E}\kern-.125emX}}

\begin{document}

\IEEEoverridecommandlockouts
\IEEEpubid{\makebox[\columnwidth]{978-1-7281-7374-0/20/\$31.00 \copyright 2020 IEEE \hfill} \hspace{\columnsep}\makebox[\columnwidth]{ }}

\title{DDoSNet: A Deep-Learning Model for \\Detecting Network Attacks %in SDN Network
}

\author{
Mahmoud~Said~Elsayed$^{1}$,
Nhien-An~Le-Khac$^{1}$,
Soumyabrata~Dev$^{1,2}$,
and~Anca~Delia~Jurcut$^{1}$% <-this % stops a space
\\
$^{1}$School of Computer Science, University College Dublin, Dublin, Ireland\\
$^{2}$ADAPT SFI Research Centre, Dublin, Ireland
\thanks{The  ADAPT  Centre  for  Digital  Content  Technology  is  funded  under  the  SFI Research Centres Programme (Grant 13/RC/2106) and is co-funded under the European Regional Development Fund.}
\thanks{Send correspondence to A.\ D.\ Jurcut, E-mail: anca.jurcut@ucd.ie.
}% <-this % stops a space 
}

\maketitle

\makeatletter
\def\runningfoot{\def\@runningfoot{}}
\def\firstfoot{\def\@firstfoot{}}
\makeatother

% \and
% \and
% \and
% \and
% \and

\begin{abstract}
Software-Defined Networking (SDN) is an emerging paradigm, which evolved in recent years to address the weaknesses in traditional networks. The significant feature of
the SDN, which is achieved by disassociating the control plane from the data plane, facilitates network management and allows the network to be efficiently programmable. However, the new architecture can be susceptible to several attacks that lead to resource exhaustion and prevent the SDN controller from supporting legitimate users. One of these attacks, which nowadays is growing significantly, is the Distributed Denial of Service (DDoS) attack. DDoS attack has a high impact on crashing the network resources, making the target servers unable to support the valid users. The current methods deploy Machine Learning (ML) for intrusion detection against DDoS attacks in the SDN network using the standard datasets. However, these methods suffer several drawbacks, and the used datasets do not contain the most recent attack patterns - hence, lacking in attack diversity. 

In this paper, we propose DDoSNet, an intrusion detection system against DDoS attacks in SDN environments. Our method is based on Deep Learning (DL) technique, combining the Recurrent Neural Network (RNN) with autoencoder. We evaluate our model using the newly released dataset CICDDoS2019, which contains a comprehensive variety of DDoS attacks and addresses the gaps of the existing current datasets. We obtain a significant improvement in attack detection,
as compared to other benchmarking methods. Hence, our model provides great confidence in securing these networks.
\end{abstract}


%
% Keywords. The author(s) should pick words that accurately describe the work being
% presented. Separate the keywords with commas.
%
% A "teaser" image appears between the author and affiliation information and the body 
% of the document, and typically spans the page. 

%
% This command processes the author and affiliation and title information and builds
% the first part of the formatted document.
\maketitle

\section{Introduction}
 
Software-Defined Networking (SDN) is a new technology that facilitates management and programmability of the network system. SDN makes the network more reliable by centralizing it through separating the control plane from the data plane. However, the emerging paradigm is subjected to many security vulnerabilities and new faults that can be used by attackers to create different types of malicious attacks~\cite{kreutz2013towards, kreutz2014software}. Further, the common threats and attacks which can exploit the classical network infrastructure can also exist in the SDN environment~\cite{scott2015survey}. Moreover, these attacks can impact the whole SDN system that includes multi-devices from different vendors, unlike in the traditional network where in general an attack, mainly crashes a part of the network devices from a single vendor only without affecting the entire network. There are many attack vectors that can exploit the SDN network~\cite{yoon2017flow}. 

One of the most common and dangerous types of attacks is Distributed Denial of Service (DDoS) attack, which can prevent legitimate users from access their network services. DDoS attack can deplete the network resources or target the servers by flooding the network with a large number of volume traffics. In addition, because of the IoT era,  there are many devices that can be connected to the Internet. Hence, attackers can exploit many types of DDoS attacks by leveraging massive numbers of bots from different locations. The execution of DDoS attacks using bots devices is hard to discover. Additionally, these attacks consume the network resources in a very short time. 
S. Mohammed~\cite{mohammed2018new} reported that an heavy DDoS attack can cause a loss reaching $\$ 100,000$ per hour for some organizations along with damaging the trust of its customers. The DDoS attacks can overwhelm different layers of SDN, such as the communication channels between the SDN controller and open flow switches or the channel between the controller and application layer. Since SDN has a single point of failure, if it is overwhelmed by any DDoS attack, then the whole network simultaneously will fail.

There are two main types of DDoS attacks: volumetric attack~\cite{talpur2016survey} and attack on the application layer~\cite{yevsieieva2017analysis}. The volumetric attack or flooding attack saturates and consumes the bandwidth of the network infrastructure. It commonly uses layer 3 or layer 4 protocols to generate high volumes of traffic, and common types include ICMP, UDP, and TCP-SYN flood. The application Layer attack is more sophisticated and in most cases uses  less bandwidth for starting.  It targets specific applications or services and slowly exhausts the network resources. Attackers can keep the connection open as long as possible by sending the requested data with a very small packet window. Examples include HTTP and DNS attacks. 

In recent years, a large number of approaches using Machine Learning (ML) techniques were proposed to detect DDoS attacks~\cite{yang2018ddos, santosmachine, elsayed19, kokila2014ddos}. The majority of those studies employed the ML for feature selection methodology to get high performance from the classifiers systems. However, there are several problems from applying ML on feature selection techniques. First, with the success of big data and IoT technology, the size of network traffics are growing fast. The classical ML classifications have difficulties to work with a large amount of data due to its limited ability in feature learning.  Classical ML gives better results to find similarities in known attacks than discovering the outliers activities for unknown malicious attacks~\cite{sommer2010outside}.  In addition, the false alarm is significantly high when the amount of data becomes extremely large. Moreover, the feature selection in classical techniques mainly needs experts, as the choice of the proper features is achieved manually in these methods.

Recently, Deep Learning (DL) achieved great success in many different applications, such as face recognition, image processing, and natural language  translation. DL has the ability to extract the raw features from the data  without any human intervention. It can satisfy high performance rate by finding the correlation on raw data automatically. Hence, with the advent of DL based  models,  the  accuracy  in  detecting  attacks  has  further improved. However, the temporal correlations of network traffic often generate time-series data~\cite{tang2018deep}, and training the simplest form of DL algorithms with sequential traffic can cause a loss in some data information. To cover this problem and avoid any loss, we use Recurrent Neural Network (RNN) technique for our proposed solution. The RNN considers the previous computations with the current events at any input stage. Training the model with such methods can keep all data information with minimum loss. We use standard RNN as we do not focus on learning long-term temporal dependencies. The standard RNN is simple and takes less time for training while comparing to the different RNN techniques~\cite{prabowo2018lstm}.
In this paper, we propose DDoSNet: a deep learning technique based on RNN-autoencoder for the detection of DDoS attacks on the SDN. The proposed model has the best performance in terms of precision, recall, F1-score and accuracy compared to different classical techniques.

The contribution of this paper includes the following:

\begin{itemize}

\item We leverage and propose a deep learning approach based on RNN-autoencoder for detection of DDoS attacks on the SDN (DDoSNet). We combine RNN-autoencoder with  softmax regression model at the output layer to classify the network traffic into malicious or normal.
 
\item We evaluate our model using the new released dataset CICDDoS2019, which contain a comprehensive variety of DDoS attacks and addresses the gaps of the existing current datasets. 
 
\item We benchmark several state-of-the-art ML models that are well known for detection of DDoS attacks  and we evaluate our proposed model in terms of  precision, recall, F1-score and accuracy.  Our proposed  method  has  the  best  performance.

\end{itemize}

The rest of the paper is structured as follows: Section~\ref{sec:rwork} discusses related work on detection of DDoS attacks in SDN environment. Section~\ref{sec:proposed} introduces our proposed technique. The new dataset and the evaluation of our proposed model are presented in Section~\ref{sec:evaluation}. Finally, Section~\ref{sec:conc} concludes the paper and presents some future work.

\section{Related Work}
\label{sec:rwork}
In this section, we briefly discuss the most recent and popular mechanisms that have been used for the detection of DDoS attacks in SDN environments. 

Ye \textit{et al.} in~\cite{ye2018ddos} used Support Vector Machine (SVM) classification algorithm to detect DDoS attack in the SDN network. The authors adopted six tuple features that can be collected from  SDN controller for the training stage. The dataset samples are collected by simulating the SDN network using mininet and flood controller with five virtual hosts. Three different DDoS scenarios are generated during simulation phase, including UDP, TCP SYN and ICMP flood traffic. Rahman \textit{et al.} in~\cite{rahman2019ddos} conducted four different ML techniques to detect DDoS attacks under the context of SDN. The authors simulated the normal and malicious traffic to create Training and Testing Dataset. Two DDoS samples (TCP and ICMP floods) are generated using hping3 tool. The experiment results showed that the J48 has better accuracy in comparison to the other evaluated techniques. Abhiroop \textit{et al.} in~\cite{abhiroop2018machine} used three different ML algorithms: SVM, Naive Bayes (NB), and Neural Network (NN) to detect flow-table overflow attack inside the SDN data plane. The author's employed open flow protocol to collect the tuple features from open flow switches to create training data. Scapy tool is used to generate three flood traffics such as TCP, UDP and ICMP flood. Five features are adopted for ML techniques, and the experiment results show that the SVM gives a low accuracy rate compared to the last two classifiers.  

The authors in~\cite{karan2018detection} introduced two detection models for DDoS attacks against SDN network. In the first stage, the signature-based snort detection system was used to collect network traffic. In the last stage, SVM and Deep Neural Network (DNN) techniques are employed for attack classification. The authors used the KDDCUP’99 dataset for training the two detection modules based on 7 features from a total of 41. The experiment results proved that the DNN performs better than SVM with accuracy rate of $92.30$  and $74.30$\%, respectively. Mohammed \textit{et al.}~\cite{mohammed2018new} proposed a new framework for detection of DDoS attacks on SDN. NSL-KDD dataset with 25 selected features was used for training NB classifier. The authors run three different selection algorithms (Genetic, Ranker, Greedy) together and select the combined features from the dataset. The average values of Precision, Recall, and F1-score are $0.81$, $0.77$, and $0.77$, respectively.

Most of the detection systems in the literature, which simulated the SDN network to generate the DDoS attacks dataset consider only a reduced number of malicious activities, only for IP or TCP protocol, without considering any applications layer DDoS attacks. One of the challenges to detect application layer DDoS attacks is the high similarity of attacks and the benign behaviors. Hence, there is a lack of available features to define such attacks, and therefore, many detection systems are not suitable to identify it~\cite{yadav2016detection}. In addition, the simulated traffics are generated using tools like Scapy and Hping3. Hence, the produced dataset is small and does not contain the complete traffic to provide accurate results. Additionally, the existing techniques that are using public datasets to train anomaly detection systems suffer of several issues. For example, most of the datasets are out of date and do not contain the new type of attacks traffic. Besides, they contain few types of attacks to address all trends found in the Internet today. A comprehensive and valid dataset has a great impact on the evaluation of detection algorithms and techniques systems. 
Thus, to evaluate our proposed model, we are using the newest public available dataset CICDDoS2019~\cite{Iman2019DDoS}, which  contain  a  comprehensive  variety of  DDoS  attacks  and  addresses  the  gaps  of  the  existing current datasets.

\section{Proposed Model}
\label{sec:proposed}
In this section, we introduce the architecture of our proposed model DDoSNet. DDoSNet is  based on autoencoder and RNN deep neural network.

Autoencoder is a type of artificial natural network, which is deployed in many applications such as data noise filtration or image processing. %It has a power discriminatory for feature extraction from input data.
The reason we used autoencoder in this work is because of the fact that the autoencoder can significantly increase the anomaly detection accuracy compared to linear and kernel Principal Component Analysis (PCA)~\cite{sakurada2014anomaly}. It can detect subtle anomalies, which linear PCA fails to detect. Furthermore, the autoencoder is easy to train and does not require complex computation like kernel PCA.

The autoencoder is composed of three separate layers. The first layer is the input layer, which receives the original input $\mathbf{X}_i$ and encodes, decodes it through several consecutive hidden layers (encoder and decoder blocks). The encoded features at the encoder phase have a smaller dimension than the input data, while in the decoder phase, the encoded features are reconstructed in reverse order to generate the final output at last layer. The generated output feature vector  $\widehat{\mathbf{X}_i}$ is approximately equal to the original input data. We combined the autoencoder with the standard RNN mechanism to generate a better detection model for detection of DDoS attacks. RNN can address the problem of tradition feed-forward neural networks~\cite{yin2017deep}. As a result, it can create much powerful models with high classification accuracy. RNN is widely applied in different domain applications such as language processing and speech recognition. Unlike the feed-forward neural networks, the cyclic connections of the RNN can be effectively used for modeling sequences~\cite{jiang2018deep}. In RNN, for the given input vector sequence \(X=(x_{1}, x_{2}, x_{3},...., x_{t})\), we can compute the hidden vector \(Z=(z_{1}, z_{2}, z_{3},...., z_{t})\) and  output vector sequence \(F=(f_{1}, f_{2}, f_{3},...., f_{t})\) at time t using Eq.\ref{eq1} and Eq.\ref{eq2}, respectively.

\begin{equation}
z_{t}=\sigma (W_{xz}X_{t}+W_{zz}z_{t-1}+b_{h}),
    \label{eq1}
\end{equation}

\begin{equation}
f_{t}=W_{zf}h_{t}+b_{f},
    \label{eq2}
\end{equation}

where $\sigma$ is the activation function, $W$ is the weight, $b$ is the bias and $z_{t-1}$ is the state at time $t-1$.

\begin{figure}[htb]
  \begin{center} \includegraphics[width=0.45\textwidth]{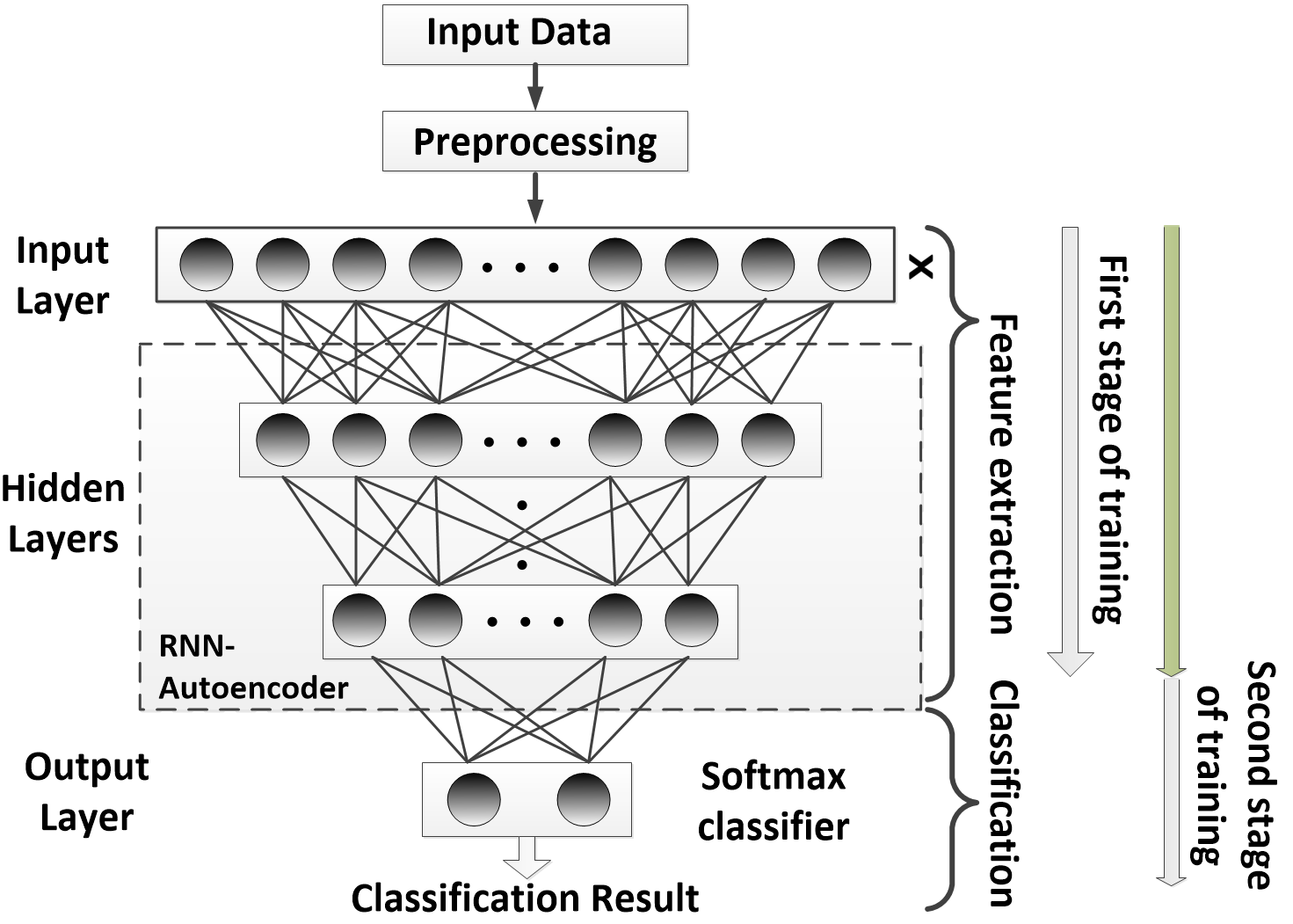}
     \end{center}
  \caption{Our proposed model for detecting attacks in a SDN network.}
 \label{fig:pro:archite}
\end{figure}

Our model can learn a vector representation of the input data in a comprehensive manner. Figure~\ref{fig:pro:archite} describes our proposed architecture for intrusion detection. The proposed model contains two stages: unsupervised pre-training stage and fine-tuning stage. The first stage is employed to extract the useful feature representation of the input data. We trained the RNN-autoencoder architecture in an unsupervised setting, without the labels to generate the compressed representation of the inputs data. Each layer of the autoencoder is a simple RNN layer. We used four hidden layers in the encoder phase, with the number of channels equal to $64$, $32$, $16$, and $8$, respectively. The channel number of the decoder phase is in the reverse order of the encoder block with number of channels as $8$, $16$, $32$, and $64$, respectively. After we obtain the desirable values of the weight and bias, the RNN-autoencoder is able to learn the hierarchical features from the unlabeled data. Next, fine-tuning training (supervised) is conducted to train the last layer of the network using labeled samples. Implementing the fine-tuning using supervised training criterion can further optimize the whole network\cite{bengio2007greedy}. 
We use \texttt{softmax} regression layer with two channels at the top layer. The output of the \texttt{softmax} function is in the range of ($0$, $1$) for each label class, where the total probability values for all classes is equal to unity.

\begin{figure*}[htb]
  \begin{center} \includegraphics[width=0.75\textwidth]{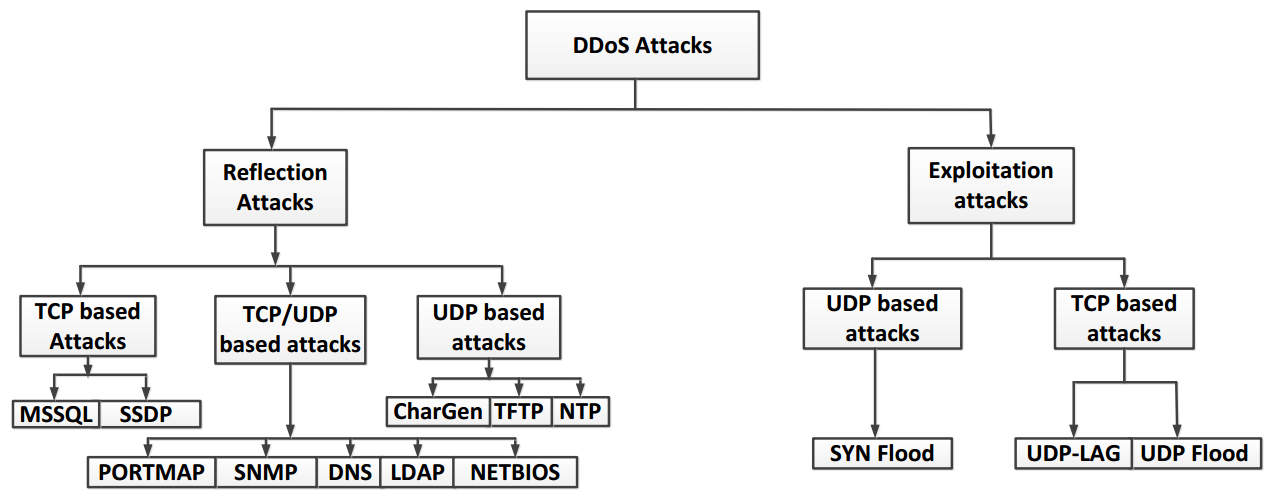}
  \end{center}
  \caption{The distribution of DoS and DDoS attack inside the CICDDoS2019 dataset~\cite{Iman2019DDoS}.}
  \label{fig:dataset-table}
\end{figure*}

\section{Evaluation Methodology}
\label{sec:evaluation}

\subsection{Dataset}
One of the most extensive challenges for ML/DL intrusion detection approaches is the availability of the datasets. The main reason for the lack of datasets in the intrusion detection domain returns to privacy and illegal issues. The network traffic contains very sensitive information, where the availability of such information can reveal customers and business secrets, or even the personal communication. To cover the previous gap, many researchers simulate their own data to avoid any sensitive concerns. However in these situations most of the datasets generated are not comprehensive and the row samples considered are not sufficient to cover the application behaviors.  The most popular public datasets, which have been used extensively for intrusion detection are KDDCUP’99~\cite{divekar2018benchmarking}, NSL-KDD~\cite{tavallaee2009detailed}, Kyoto 2006+~\cite{song2006description}, ISCX2012~\cite{shiravi2012toward}, and CICIDS2017~\cite{sharafaldin2018toward}. More details about different datasets used for intrusion domain can be found in~\cite{ring2019survey}.
 
In this paper, we evaluate our proposed classifier using the new released  CICDDoS2019 dataset~\cite{Iman2019DDoS}. The dataset contains a large amount of different DDoS attacks that can be carried out through application layer protocols using TCP/UDP. The taxonomy of  attacks in the dataset are performed in terms of exploitation-based and reflection-based attacks. The dataset was collected in two separated days for training and testing evaluation. The training set was captured on January 12\textsuperscript{th}, 2019, and contains 12 different kinds of DDoS attacks, each attack type in a separated PCAP file. The attack types in the training day includes UDP, SNMP, NetBIOS, LDAP, TFTP, NTP, SYN, WebDDoS, MSSQL, UDP-Lag, DNS, and SSDP DDoS based attack. The testing data was created on March 11\textsuperscript{th}, 2019, and contains 7 DDoS attacks SYN, MSSQL, UDP-Lag, LDAP, UDP, PortScan, and NetBIOS. The distribution of the different attacks in the dataset is shown in Figure~\ref{fig:dataset-table}. The PortScan attack in the testing set not present in the training data for intrinsic evaluation of detection system. The dataset contains more than 80 flow features\footnote{\url{http://www.netflowmeter.ca/netflowmeter.html}.}  and was extracted using CICFlowMeter tools\cite{ draper2016characterization}. 
%An overview of subset features within the CICDDoS2019 dataset is described in Table~\ref{table:features}.
%Most of the relevant extracted features are described in Table~\ref{table:features}.
%The list of all the features and its corresponding descriptions are mentioned in~\cite{CICFlowMeter-2017}.
The CICDDoS2019 dataset is available on the website of Canadian Institute for Cybersecurity\footnote{\url{https://www.unb.ca/cic/datasets/ddos-2019.html}.} in both PCAP file and flow format based.

\begin{comment}
\begin{table}[htb]
\caption{Some Features in CICDDoS2019 dataset}
\centering
\normalsize 
\begin{tabular}{l|l}
\textbf{Feature name} & \textbf{Description} \\
\hline 
fl\_dur   & Duration of the flow in Microsecond \\
tot\_fw\_pkt  & Total packets in the forward direction  \\
tot\_bwd\_pkt  & Total packets in the backward direction \\
fl\_pkt/s  & Number of flow packets per second \\
fl\_byte/s  & Number of flow bytes per second  \\
fwd\_pkt/s  & Number of forward packets per second \\
bwd\_pkt/s  &  Number of backward packets per second \\
min\_pkt\_len   & Number of backward packets per second \\
max\_pkt\_len  & Maximum length of a packet \\
pkt\_len\_mean   & Mean length of a packet \\
syn\_f\_cnt  & Number of packets with SYN \\
avg\_pck\_sz  & Average size of packet \\
rst\_f\_cnt  & Number of packets with RST \\
\hline 
\end{tabular}
\label{table:features}
\end{table}
\end{comment}

\begin{figure}[htb]
\centering
  \begin{center} \includegraphics[width=0.35\textwidth]{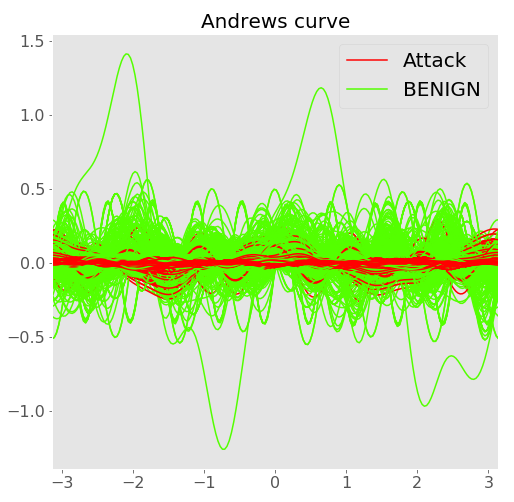}
  \end{center}
  \caption{We plot the Andrews curve for the CICDDoS2019   dataset. It is clear that the feature space obtains high degree of non-linearity.}
  \label{fig:and-curve}
\end{figure}

The classical ML techniques are  working better in feature extraction instead of raw data. Selecting the proper features in intrusion systems is not an easy task and requires the help of experts. Furthermore, the attack scenarios evolve every day, and thus selecting relevant features for a specific type of attack is not a suitable solution.     
The majority of the real-world data take the approach of non-linear or multivariate data. It is hard to visualize the data in more than three-dimension. We use the Andrew curve~\cite{spencer2003investigating} for better understanding of the distribution of multidimensional data. Andrew curve is based on the Fourier series, and it is widely applied for multivariate data visualization. To analyze the testing data using the Andrew curve, we select random 10 \%  of sample data and reduce its feature dimensions to 20 using the PCA technique to get optimal results and speed the calculation. The graph in Figure~\ref{fig:and-curve} shows the complex relationship for malicious and normal features in the dataset. Each line in the curve represents different observations in the data -- \emph{Attack} and \emph{Benign}. It is clear that the subgroups of input data are interlinked with each other, which indicates a high degree of non-linearity in the feature space of observed data. As a result, it shows that the classical ML methods are not operating well with multivariate datasets.

%\subsection{Loss trend of our proposed model}

\subsection{Data Preprocessing}
We prepare the data to be suitable for the training model directly. CICDDoS2019 dataset  is available in a flow-based format where more than 80 features are extracted using CICFlowMeter. We follow a few steps to prepare the data before the module training. 
\begin{itemize}
\item Removing socket features: we remove all of the socket features like source and destination IP, Source and destination port, timestamp, and flow ID. These features vary from network to network, and we need to train the model with packet characteristics itself. Furthermore, both the intruder and normal users can have the same IP address. Therefore, training the DL model with socket information can cause an overfitting problem, as the model can be biased to the socket information. We obtained a  number of $77$ features for the model input after removing the unwanted features.
 \item Cleaning the data: the original data contains a large amount of missing (nan) and infinity values - we remove all these values from the data.
 \item  Normalize the input data: the features data have different numerical values. Training the model directly with the original data can cause classification error, and then the model takes much time during its training. We normalize the data where the minimum value is zero, and the maximum is one.
 \item  Encoding the labeled data: we trained our model for binary classification to classify the input traffic into normal or malicious. Therefore, we consider all DDoS classes as attack category, besides the normal traffic. Then, we are encoding the string value for normal and attack label to binary value of 0 and 1, respectively.
   \end{itemize}

\subsection{Experimental Setup}
\subsubsection{Training RNN-autoencoder}

%To evaluate the model performance and avoid the overfitting problem, 
We split the dataset into three sections of training, validation, and testing subset. We build the model using the training set to adjust the weight on the neural network. The validation set is used to fine-tune the experiment parameters \textit{i.e.} classifier architecture (not weights) like the number of hidden layers in the proposed model. Besides, the test set is used to estimate the model accuracy or performance. In this paper, we used the train-test split technique for the model evaluation instead of the k-fold cross-validation technique. Although cross-validation is a widely used method in classification evaluation, we do not use it in our case. The inherent serial correlation of the time series data makes the use of cross-validation less suitable for these problems~\cite{bergmeir2018note}.% seem to error as it does not suitable K-fold CV takes a large time to fit the model i.e. K-fold CV takes k-times more than train-test split. Further, we deal with the network traffic as a time series data. 

\begin{figure}[htb]
  \begin{center}
    \includegraphics[width=0.35\textwidth]{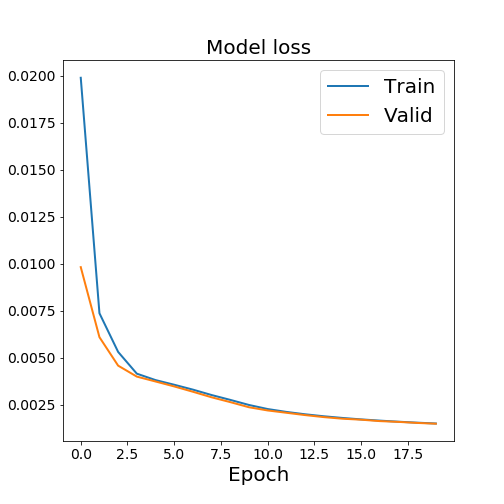}
  \end{center}
  \caption{Trend of training and validation loss over the number of epochs.}
  \label{fig:loss}
\end{figure}

In this experiment, the \texttt{softmax} layer takes the decoder output and classifies the input data into normal or attack traffic. We used \texttt{categorical-crossentropy} as loss function with \texttt{adam} optimizer and \texttt{ReLU} function for activation in all different layers. We trained the model using the number of epochs of $50$ and a batch size of $32$. Figure~\ref{fig:loss} describes the trend of training and validation loss over the number of epochs. We observe that the training and validation loss converges and reaches the minimum value after $10$ epochs. We chose the model with the least validation loss. We run several experiments using different values of learning rate to get optimal results. The selective learning rate for this experiment is $0.0001$. More details about the impact of  the learning rate are discussed subsequently. 

\subsubsection{Hyper-parameter Tuning}

The trained model behavior is directly controlled by the values of the hyper-parameters, where selecting the best values plays a key role in the success of neural network architecture. However, selecting the best values of hyper-parameters is still dependent on the best practice or human knowledge. We conducted various experiments to select the optimal values of experiment hyper-parameters. We tested the performance of the model using different learning rate values \textit{i.e.} $0.1$, $0.01$, $0.001$ and $0.0001$. Table~\ref{table:LearningRate} shows the performance metrics of the RNN-autoencoder model with different learning rates. Based on the obtained results, we can conclude that the model consumed a long time during the training process when we reduced the learning rate value. However, the model achieved batter results using a smaller value of learning rate. It can be observed that the model achieved overall accuracy up to 99\% when the learning rate is decreased to $0.0001$.
Furthermore, we changed the number of hidden layers, iteration, number of channels per hidden layer, and the activation function for each learning rate value. The best performance is achieved when we used four hidden layers. When the number of hidden layers is increased, the model accuracy remains constant, but the training increases considerably. Therefore, we use four layers in our proposed framework. As a result, the four layers are more convincing to give reasonable results.

\subsubsection{Data Partitioning}

The performance of the classification methods not only depends on the used technique but also on the manner in which training- and testing- data is partitioned. %depends on the training and testing data that are used. 
%Some of the previous works presented how to divide the training and testing data to achieve a high detection rate. The authors in
Previous work by Rasool \emph{et al.}\cite{rasool2019cyberpulse} tested the effect of data partitioning strategies on the accuracy of the classifier. They approved that classifier performance is increased when the training data is gradually increased. The best performance is obtained when 70\% of the input data is used for training. %It can be noticed that the classifier accuracy is significantly increased when the training data increased to 70 \%.
After we prepared the input data, the final training data has $77$ input features. Furthermore, the distribution of samples in the dataset is different and extremely large. We use samples from each attack type to obtain a balanced dataset with respect to different types of attacks. %get equivalent distributed data and make the average accuracy very close for each attack type. 
In our case, the total number of samples for training and validation sets are $161523$ and $46150$, respectively. To get a realist detection rate, we used attack records in the testing set that are not represented in the training phase. The total number of records in testing set is $23000$ samples.

\begin{table}[htb]
\centering
\caption{Performance evaluation of the proposed model with different learning rate.}
\resizebox{8.9cm}{!}{
\begin{tabular}{|c|c|c|c|c|c|c|c|}
\hline
\multirow{3}{*}{\textbf{Learning rates}} & \multicolumn{7}{c|}{\textbf{Performance metrics}} \\ \cline{2-8} 
 & \multicolumn{2}{c|}{Precision} & \multicolumn{2}{c|}{Recall} & \multicolumn{2}{c|}{F-score} & \multirow{2}{*}{Accuracy (\%)} \\ \cline{2-7}
 & Attack & Benign & Attack & Benign & Attack & Benign &  \\ \hline
0.1 & 0.97  & 0.97 & 0.96 & 0.97 & 0.96 & 0.97  & 97 \\ \hline
0.01 & 0.98 & 0.99 & 0.99 & 0.98  & 0.98 & 0.98 & 98 \\ \hline
0.001   & 0.98  & 0.99  & 0.99 & 0.98 & 0.99 & 0.99 & 99 \\ \hline
0.0001  & 0.99 & 1.00 & 0.99 & 0.99 & 0.99 & 0.99  & 99 \\ \hline
\end{tabular}
\label{table:LearningRate}
}
\end{table}

\subsection{Evaluation}
%\subsubsection{Classification Metrics}

%but SVM takes more than 5 hours during its training.

\begin{figure}[htb]
  \begin{center}
    \includegraphics[width=0.3\textwidth]{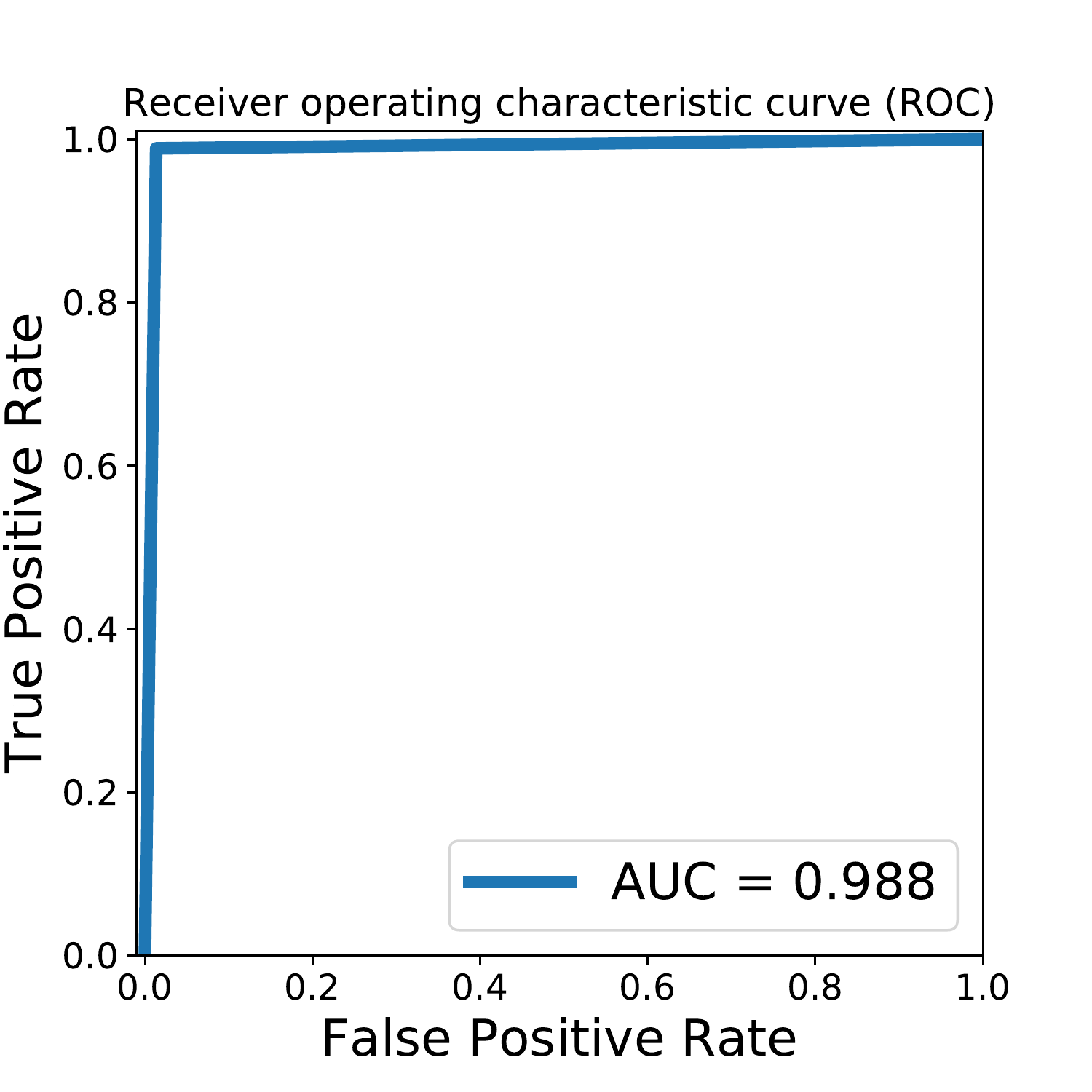}
  \end{center}
  \caption{Receiver Operating Curve (ROC) for our proposed model.}
  \label{fig:roc}
\end{figure}

%\subsubsection{Receiver Operating Characteristic (ROC)}
We use the Receiver Operating Characteristic (ROC) curve to evaluate how well the model performs accurately. The ROC curve indicates the relation between two parameters: True and False classes. The area underneath the ROC Curve (AUC) measures separability between false positive and true positive rate. Figure~\ref{fig:roc} shows that our model gives AUC of $98.8$, which means that our proposed model is able to separate $98.8\%$  of positive and negative classes successfully. 

%\subsubsection{Confusion Matrix}
We also illustrate the confusion matrix in order to describe the classification performance of our model. The confusion matrix summarizes the correct and false predictions. Table~\ref{table:cf-matrix} indicates that our model can detect all the attack and normal classes with good accuracy of $0.99$.

\begin{table}[htb]
\centering 
\caption{Confusion matrix on the four distinct events (TP, FP, TN, FN) obtained using our proposed approach.}
\begin{tabular}{c|c|c}
\textbf{Attack} & 0.99 & 0.01 \\ \hline
\textbf{Normal} & 0.01 & 0.99 \\ \hline
 & \textbf{Attack} & \textbf{Normal} \\ 
\end{tabular}
\label{table:cf-matrix}
\end{table}

Furthermore, we also use various metrics to evaluate our proposed model, such as precision, recall, precision, F-score and accuracy, in order to have a systematic benchmarking analysis with other related approaches. These metrics are commonly used in intrusion detection systems and are defined as follows: 

\begin{equation}
	\mbox{Precision}=\frac{TP}{TP+FP} 
    \label{eq3}
\end{equation}   

\begin{equation}
	\mbox{Recall}=\frac{TP}{TP+FN}
    \label{eq4}
\end{equation}   

\begin{equation}
	\mbox{F-score}=\frac{2\times\mbox{Precision}\times\mbox{Recall}}{\mbox{Precision}+\mbox{Recall}}
    \label{eq5}
\end{equation}

\begin{equation}
	\mbox{Accuracy}=\frac{TP + TN}{TP + TN + FP + FN}
    \label{eq6}
\end{equation}

where True Positive (TP) and True Negative (TN) represent the values that are correctly predicted. In contrast, False Positives (FP) and False Negatives (FN) indicate misclassified  events.

We compare our model with various classical ML techniques. In this evaluation, we consider 6 different algorithms Decision Tree (DT), NB, Booster, Random Forest (RF), SVM, and Logistic Regression (LR). Table~\ref{table:result} represents the classification metrics for normal and attack classes of our model with different classical techniques. The obtained results show that the DDoSNet model performs better than the other ML algorithms. We observed that our proposed model performed the best, followed by LR and SVM. The NB classifier performed poorly, primarily because the NB assumed that all attributes are independent of each other~\cite{rennie2003tackling}. Therefore, its performance got impacted, as the considered feature attributes are dependent on each other.  %the performance accuracy are affected when the feature attributes are dependent on each other

\begin{table}[htb]
\centering
\caption{Performance evaluation of the proposed model with other classical ML algorithms. Our proposed approach performs best, as compared to the other benchmarking algorithms.}
\resizebox{8.9cm}{!}{
\begin{tabular}{|c|c|c|c|c|c|c|c|}
\hline
\multirow{2}{*}{\textbf{Techniques}} & \multicolumn{2}{c|}{\textbf{Precision}} & \multicolumn{2}{c|}{\textbf{Recall}} & \multicolumn{2}{c|}{\textbf{F-score}} & \multirow{2}{*}{\textbf{Accuracy (\%)}} \\ \cline{2-7}
 & Attack & Benign & Attack & Benign & Attack & Benign &  \\ \hline
NB  & 1.00  & 0.53  & 0.17  & 1.00 & 0.29  & 0.69 & 57 \\ \hline
DT & 0.70 & 0.98  & 0.99  & 0.54 & 0.82 & 0.70 & 77  \\ \hline
Booster & 0.76  & 0.99& 0.99& 0.67  & 0.86& 0.80  & 84  \\ \hline
RF & 1.00   & 0.78 & 0.74 & 1.00  & 0.85 & 0.88 & 86  \\ \hline
SVM & 0.99 & 0.88 & 0.88 & 0.99 & 0.93 & 0.93 & 93  \\ \hline
LR  & 0.93   & 0.99  & 0.99 & 0.91  & 0.96  & 0.95 & 95 \\ \hline
\textbf{Proposed DDoSNet}  & \textbf{0.99}  & \textbf{1.00}    & \textbf{0.99} & \textbf{0.99} & \textbf{0.99}  & \textbf{0.99} & \textbf{99}  \\ \hline
\end{tabular}

\label{table:result}
}
\end{table}

\subsection{Discussions}

%\textcolor{red}{

The novel contribution of this paper is to represent the potential of DL for anomaly detection systems. We proposed a new DL technique based on RNN-Autoencoder to classify the input traffic into normal or malicious types. The proposed DL model can reduce the data dimensionality by automatically extracting the %high level of abstract 
features from input data. Compared to shallow learners, our technique achieved the best performance in terms of precision, recall, F1-score, and accuracy. The ability of DL to deal with a high degree of complex non-linear relationship makes them promising techniques for detecting network intrusion. It can be used to tackle the limitation of the traditional classification methods, which are implemented to identify the anomalies traffic based on domain knowledge~\cite{liu2019cnn}. However, with the continuous development of Internet traffic and the advent of the era of big data, many new attack characteristics are derived from known attacks, and it is hard to distinguish between them. Our model can be implemented as an application layer of the SDN controller, such as in~\cite{alshamrani2017defense}. The detected malicious traffic can be blocked using null routing method or redirected to a honeypot server for further investigation.

\section{Conclusion and Future Work}
\label{sec:conc}
Network virtualization leads to new threats and new exploitable attacks that the ones already existing in the traditional network. The DDoS attack class is considered one of the most aggressive attack types in recent years, causing a critical impact on the whole network system. In this paper, we proposed a new model DDoSNet that is based on DL for the detection of DDoS attacks against SDN network. We used the new released CICDDoS2019 dataset for training and evaluation of our proposed model. The dataset contains comprehensive and most recent DDoS types of attacks.  The evaluation of our model showed that DDoSNet gives the highest evaluation metrics in terms of recall, precision, F-score, and accuracy compared to the existing well known classical ML techniques. 

In the future, we are interested in testing the performance of our proposed model on other datasets. In this current work, we used a binary classification framework to classify the input traffic into normal and malicious types. However, it is also necessary to classify each individual attack types separately. We intend to extend our work to a multi-class classification framework. Furthermore, we will simulate the SDN network with various types of environments and attack traffics to create a heterogeneous dataset that can effectively represent the current internet traffic.

\balance

\balance

\end{document}